\documentclass[]{spie}  

\usepackage{amsmath,amsfonts,amssymb}
\usepackage{graphicx}
\usepackage[colorlinks=true, allcolors=blue]{hyperref}

\usepackage[dvipsnames]{xcolor}
\usepackage{color}
\usepackage{xcolor}

\title{Flux reconstruction for the NIR camera CAGIRE at the focus of the Colibrí telescope}

\author[a]{Alix Nouvel de la Flèche}
\author[a]{Jean-Luc Atteia}
\author[a]{Hervé Valentin}
\author[a]{Marie Larrieu}
\author[a]{Jérémie Boy}
\author[b]{Olivier Gravrand}
\author[c]{Olivier Boulade}
\author[d]{Jean-Claude Clemens}
\author[d]{Aurélia Secroun}
\author[d]{Eric Kajfasz}
\author[d]{Olivier Llido}
\author[e]{Stéphane Basa}
\author[e]{François Dolon}
\author[e]{Johan Floriot}
\author[e]{Simona Lombardo}
\author[f]{Adrien Lamoure}
\author[f]{Laurent Rubaldo}
\author[f]{Bruno Fieque}
\author[f]{Julien Roumegoux}
\author[g]{Hervé Geoffray}
\author[h]{Alan M. Watson}
\author[h]{William H. Lee}
\author[i]{Nathaniel Butler}

\affil[a]{IRAP, Universit\'e de Toulouse, CNRS, CNES, UPS, (Toulouse), France}
\affil[b]{CEA-LETI, 17 Avenue des Martyrs, 38054, Grenoble, France}
\affil[c]{CEA-IRFU,Orme des Merisiers, Gif-sur-Yvette, France}
\affil[d]{Aix Marseille Univ, CNRS/IN2P3, CPPM, Marseille, France }
\affil[e]{Aix Marseille Univ, CNRS, CNES, LAM, Marseille, France }
\affil[f]{LYNRED, 364 Av. de Valence, Veurey-Voroize, France}
\affil[g]{CNES, 18 Av. Edouard Belin, Toulouse, France}
\affil[h]{Instituto de Astronomía, Universidad Nacional Autónoma de México, Apdo Postal 70-264, Cd. Universitaria, CDMX, México}
\affil[i]{School of Earth and Space Exploration, Arizona State University, Tempe AZ 85287, USA}

\authorinfo{Further author information: (Send correspondence to Alix Nouvel de la Flèche)\\Alix Nouvel de la Flèche: E-mail: alix.nouvel-de-la-fleche@irap.omp.eu\\  Jean-Luc Atteia: E-mail: Jean-Luc.Atteia@irap.omp.eu}

\begin{document}

\maketitle

\begin{abstract}
CAGIRE is the near infrared camera of the Colibrí robotic telescope, designed for the follow-up of SVOM alerts, mainly Gamma Ray Bursts (GRBs), and the quick imaging of sky regions where transient sources are detected by the SVOM satellite. CAGIRE is based on the Astronomical Large Format Array (ALFA) 2k x 2k SWIR sensor from the French consortium CEA-LYNRED. In the context of CAGIRE the sensor is  operated in “Up the Ramp” mode to observe the sky in a square field of view of 21.7\,arcmin on a side, in the range of wavelengths from 1.1 to 1.8\,µm. An observation with CAGIRE consists of a series of short (1-2\,minutes) exposures during which the pixels are read out every 1.3 second, continuously accumulating charges proportionally to the received flux, building a ramp. 
 
 The main challenge is to quickly process and analyse these ramps, in order to identify and study the near infrared counterparts of the bursts, within 5\,minutes of the reception of an alert. Our preprocessing, which is under development, aims at providing reliable flux maps for the astronomy pipeline. It is based on a sequence of operations. First, calibration maps are used to identify saturated pixels, and for each pixel, the usable (non saturated) range of the ramp. Then, the ramps are corrected for the electronic common mode noise, and differential ramps are constructed. Finally, the flux is calculated from the differential ramps, using a previously calibrated map of pixel non-linearities. 
 We present here the sequence of operations performed by the preprocessing, which are based on previous calibrations of the sensor response. These operations lead to the production of a flux map corrected from cosmic-rays hits, a map depicting the quality of the fit, a map of saturated pixels and a map of pixels hit by cosmic-rays, before the acquisition of the next ramp. These maps will be used by the astronomy pipeline to quickly extract the scientific results of the observations, like the identification of uncatalogued or quickly variable sources that could be GRB afterglows. 
\end{abstract}

\section{Introduction}
\label{sec:intro}

The SVOM mission (\textit{Space based multi-band astronomy Variable Objects Monitor}) aims at observing transient  high energy sources, such as gamma ray bursts (GRBs) \cite{Wei2014}.  One goal of this Sino-French mission is the study of the origin and environment of GRBs, and their use as tools to study the infancy and the evolution of the universe and its first stars.  The mission is designed to combine a space and a ground segment and to make the most of their synergy. Based on the localization of the bursts detected by ECLAIRs onboard SVOM, the ground-based robotic telescope Colibrí, operating from the  Observatorio Astronómico Nacional in San Pedro Mártir,  will quickly point the source and acquire images of the burst error box at near infrared (NIR) and visible wavelengths \cite{Basa2022}. The presence of a burst appears in the images as a new unidentified star, often with rapid luminosity variations, justifying the need for a telescope able to point the source in less than 30\,s. Moreover the visible camera does not allow to detect high redshift candidates (z $\ge$ 7.4), and the near infrared domain is needed to detect, monitor and study them. To do so, a near infrared camera, CAGIRE (CApturing Grbs InfraRed Emission), has been developed.

In this paper we describe the hereafter called preprocessing pipeline, used to reconstruct the flux received by the detector of CAGIRE, from the ramps acquired. The preprocessing pipeline has been tested on ramps taken with the RATIR instrument \cite{Butler2012}. 

\section{CAGIRE operation}
\label{sec:cagire}

\subsection{Presentation of the camera}
\label{sub:cagire}

CAGIRE is a scientific camera sensitive in the near infrared between 1.1 and 1.8\,µm, located at one of the Nasmyth foci of the robotic telescope Colibrí \cite{Basa2022}. When Colibrí receives a burst alert, it quickly points the source and CAGIRE acquires images of the sky in the NIR domain, within a square field of view of 21.7\,arcmin on a side, to cover the error boxes of ECLAIRs \cite{Godet2014}. CAGIRE acquisitions will start soon after the telescope pointing and consist of a series of short (1-2 minutes) exposures, processed right after their acquisition, in order to identify potential afterglows within 5 minutes of the reception of an alert.

\begin{figure}[ht]
        \centering
        \includegraphics[width=9.3cm]{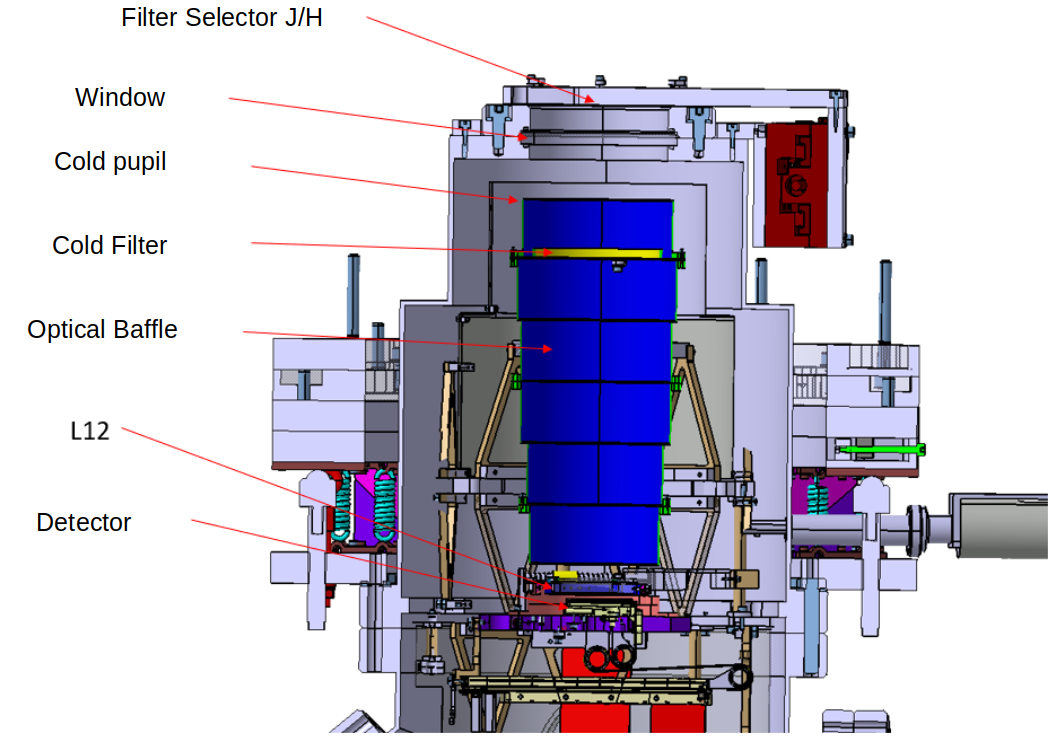}
        \caption{Model of the CAGIRE camera}
        \label{fig:cam}
    \end{figure}

The camera is composed of three subsystems: a cryostat housing the detector which is maintained at a constant temperature of 100K, a close electronics on the telescope, mounted on the structure supporting the instruments, and a remote electronics located in the control room. The light  arriving from the telescope is focused on the detector by a series of lenses in a warm optical bench. At the entrance of the cryostat it crosses a warm filter allowing to select the desired photometric channel (J or H). Then, it enters the cryostat and crosses a cold filter cutting all the wavelengths longer than 1.8\,µm, and a re-imaging cold lens, before reaching the detector (fig \ref{fig:cam}). 
    
The detector is read in science mode with 32 outputs at the readout frequency of 100\,kHz, such that the whole array is read every 1.3\,s. It will be used in up-the-ramp mode, in which the pixels continuously accumulates charges, roughly proportionally to the flux received by the camera during the exposure. Pixels are read every 1.3\,s, creating a ramp of the accumulated signal, for each pixel.  This mode allows to fulfill the requirement of having a temporal resolution better than 2\,s. The exposures will last from one to two minutes, in order to keep the sky background below few tens of percent of the sensor's dynamic range. This non-destructive acquisition mode enables more flexibility to process the data after their acquisition, and doesn't lead to data rate issues because of the localization of the telescope on ground. 
The  sky resolution of the system is of 0.64\,arcsec/pixel. This resolution implies that the detector will receive a flux from the sky background of about 150\,electrons/pixel/s for the J photometric channel and about 1250\,electrons/pixel/s for the H channel.

\subsection{The ALFA sensor}
\label{sub:alfa}

The camera CAGIRE is equipped with an Astronomical Large Format Array (ALFA) detector of 2048 by 2048 square pixels of 15\,µm on a side, manufactured by the French company LYNRED, and based on the MCT (Mercury Cadmium Telluride) technology developed at CEA-LETI \cite{Fieque2018}. The detector covers a spectral range from 0.8\,µm to 2.1\,µm, \cite{Gravrand2022} justifying the need for a blocking filter for wavelengths longer than 1.8\,µm on CAGIRE.
The detection layer in HgCdTe (MCT) is hybridized with indium bumps on a Read Out Integration Circuit (ROIC), designed and manufactured by LYNRED. 
The array is composed of 2040 by 2040 active pixels surrounded by a ring of pixels under permanent reset. Around them, reference pixels, with the same electronics as the active ones but not sensitive to light, can be used for Read-Out Integrated Circuit (ROIC) noise reduction. 
The detector is read through 32 outputs simultaneously.
More information about the detector is given in Fieque et al. 2018 \cite{Fieque2018}.

In order not to saturate the detector with the sky background, and to let enough dynamic available for bright objects, the exposures have to be shorter than 200\,s in the J band and shorter than 30-60\,s in the H band. This leads to accumulate around 30,000\,electrons from the sky background ($\sim \frac{1}{4}$ of the potential well capacity), with Poisson statistical fluctuations of approximately 170\,electrons rms per pixels. This is significantly larger than the readout noise, which is not a critical parameter for CAGIRE.

The data under study in this article have been provided by CEA-IRFU for ramps acquired with the ALFA sensor, and by CPPM for data acquired with the ROIC.

\subsection{Data processing}
\label{sub:data}

One of the advantages of CAGIRE is to be located on the ground, allowing full ramps to be stored and exploited to measure the flux received by the detector. Ideally, the signal, measured in Analog to Digital Units (ADU), should be proportional to the integrated flux, and the flux should correspond to the slope of the ramp.

However, the data are impacted by several effects due to the detector or the environment (see section \ref{sec:preproc}), requiring a pre-processing of the data before their analysis by the astronomy pipeline. Among the detector effects, conversion gain inhomogeneities (per pixel and per channel), readout noise, small fluctuations of bias voltages and pixel non-linearities affect the measure of the flux. Some non-operational pixels can also give erroneous values. Among environmental effects, cosmic-ray hits have a huge impact on the flux measurement and need to be taken into account. As CAGIRE aims at identifying quickly the NIR counterparts of high-energy transients, the ramps must be processed ``on-the-fly'', implying that one ramp must be processed before the next arrives to avoid data clogging. For this purpose, we have developed a fast and robust pre-processing pipeline, described in the next section.

\section{Preprocessing Pipeline}
\label{sec:preproc}

When CAGIRE acquires a ramp, it is necessary to quickly provide a reliable map of the flux to the astronomy pipeline. 
The preprocessing aims at constructing such maps, cleaned from instrumental effects. It is composed of several steps summarized in table \ref{tab:preproc}, and some of them require calibration maps, which are described in section \ref{sec:badpix}.
At some point, treatments depend on the number of frames in the ramp, as described in sections \ref{sub:1frame} and \ref{sub:2frame}. 
The last step, the measurement of the flux, requires differential ramps. A differential ramp is constructed by subtracting 2 consecutive frames of the ramp. It is equivalent to the ramp derivative. This process is detailed in section \ref{sec:flux}.

\subsection{Identification of saturated pixels}
\label{sub:satpix}
The first step is the identification of saturated pixels and the construction of a map of saturated pixels at the end of a ramp. This is necessary because saturation can lead to persistence, lasting for minutes to hours after the illumination\cite{Legoff2020}.

\begin{figure}[ht!]
        \centering
        \includegraphics[width=11cm]{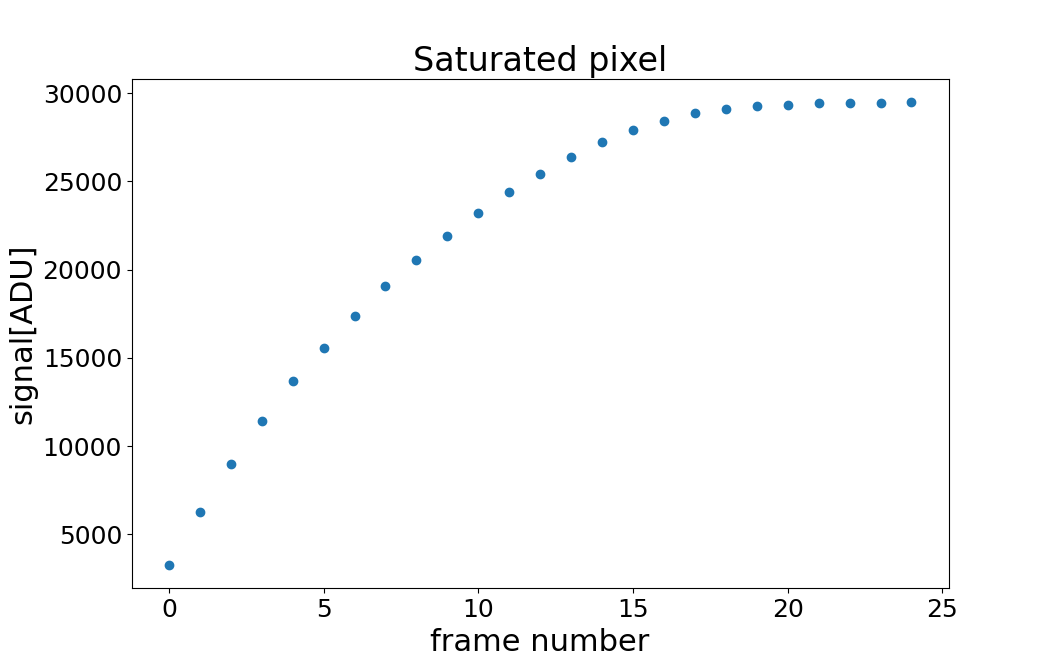}
        \caption{Example of a saturating pixel}
        \label{fig:saturating}
    \end{figure}

In order to identify saturated pixels, we compare the flux of the last frame of the ramp with the calibrated map of the saturation level. Pixels reaching the saturation level are flagged as saturated. 
The map of saturated pixels is an important piece of information, because the saturation during an acquisition can impact the following acquisitions as saturated pixels are likely to show some persistence.  This persistence can leave a detectable signal even in the absence of illumination. The map of saturated pixels thus allows to avoid confusing a GRB afterglow with the persistence of a previously saturated star. Even if saturation is not the only cause of persistence, the list of saturated pixels is an important piece of information for the correct interpretation of sources detected by the astronomy pipeline. Figure \ref{fig:saturating} shows the ramp of a saturating pixel.

\vspace{0.5cm}
    
\vspace{0.5cm}
    
\subsection{Single frame ramps}
\label{sub:1frame}
In some special cases (e.g. focus adjustment with bright sources), it may be useful to acquire ramps containing a single frame. 
In such cases the preprocessing consists of two steps: the subtraction of the electronics bias and the correction of the ROIC common mode noise using the reference pixels. This second step is described in section \ref{ssub:refpix}.

\subsection{Ramps with N \texorpdfstring{$\geq$} 2 frames}
\label{sub:2frame}

If the ramp is composed of 2 frames or more, the preprocessing involves the following steps (table \ref{tab:preproc}): the correction of the common noise using the reference pixels, the construction of a differential ramp for each pixel, the identification and correction of pixels hit by cosmic-rays, and the fit of the differential ramp by a straight line. These steps are described in some detail below.

\vspace{1cm}

\begin{table}[ht!]
    \centering
    \begin{tabular}{|l|c|l|}
  \hline
  \multicolumn{3}{|c|}{} \\ [1ex]
  \multicolumn{3}{|c|}{\textbf{Step 1: Complete ramp headers}} \\  [2ex]
  \hline
  & & \\ [-1ex]
  Ramp & Update headers & \textbf{Raw ramp}, with CAGIRE header \\
  HK information &  &  \\ [1ex]
  \hline
  \multicolumn{3}{|c|}{} \\ [1ex]
  \multicolumn{3}{|c|}{\textbf{Step 2: Find saturated pixels}} \\ [2ex]
  \hline
  & & \\ [-1ex]
  Last frame of raw ramp & Find saturated pixels & \textbf{Map of saturated Pixels} \\
  Saturation Level map &  &  \\ [1ex]
  \hline
  \multicolumn{3}{|c|}{} \\ [1ex]
  \multicolumn{3}{|c|}{\textbf{Step 3: Subtract Master Bias and correct
all frames with reference pixels }} \\ [2ex]
  \hline
  & & \\ [-1ex]
   Raw ramp & Subtract Master Bias and correct & Ramp with corrected frames \\ 
   Master Bias Map & all frames with reference pixels & \\[1ex]
  \hline
  \multicolumn{3}{|c|}{} \\ [1ex]
  \multicolumn{3}{|c|}{\textbf{Step 4: Construct Corrected Differential Ramp}} \\  [2ex] 
  \hline
  & & \\ [-1ex]
  Corrected ramp &  & Corrected differential ramp 1 (CDR1) \\ [1ex]
  \hline
  \multicolumn{3}{|c|}{} \\ [1ex]
  \multicolumn{3}{|c|}{\textbf{Step 5: Flag Cosmic Rays (CR) candidates}} \\  [2ex] 
  \hline
  & & \\ [-1ex]
  CDR 1 & CR candidates Identification & \textbf{Map of CR candidates} \\ 
        & (3 parameters) & Corrected differential ramp 2  (CDR 2)\\  [1ex]
  \hline
  \multicolumn{3}{|c|}{} \\ [1ex]
  \multicolumn{3}{|c|}{\textbf{Step 6: Flux estimation}} \\  [2ex] 
  \hline
  & & \\ [-1ex]
  CDR 2 & Flux estimation & \textbf{Flux and variance maps} \\ 
  Map of non-linearity coefficients & (1 parameter) &  \\ [1ex]
  \hline
    \end{tabular}
    \medskip
    \caption{Pre-processing pipeline for ramps with N $\ge$ 2 frames. For each step, the left column gives the input parameters, the middle column indicates the computation, and the right column the output. Files made available by the preprocessing for the astronomy pipeline are shown in bold.}
    \label{tab:preproc}
\end{table}

\subsubsection{Correction with reference pixels}
\label{ssub:refpix}

\begin{figure}[ht!]
\begin{minipage}[c]{0.42\linewidth}
    \begin{center}
             \includegraphics[width=0.85\columnwidth]{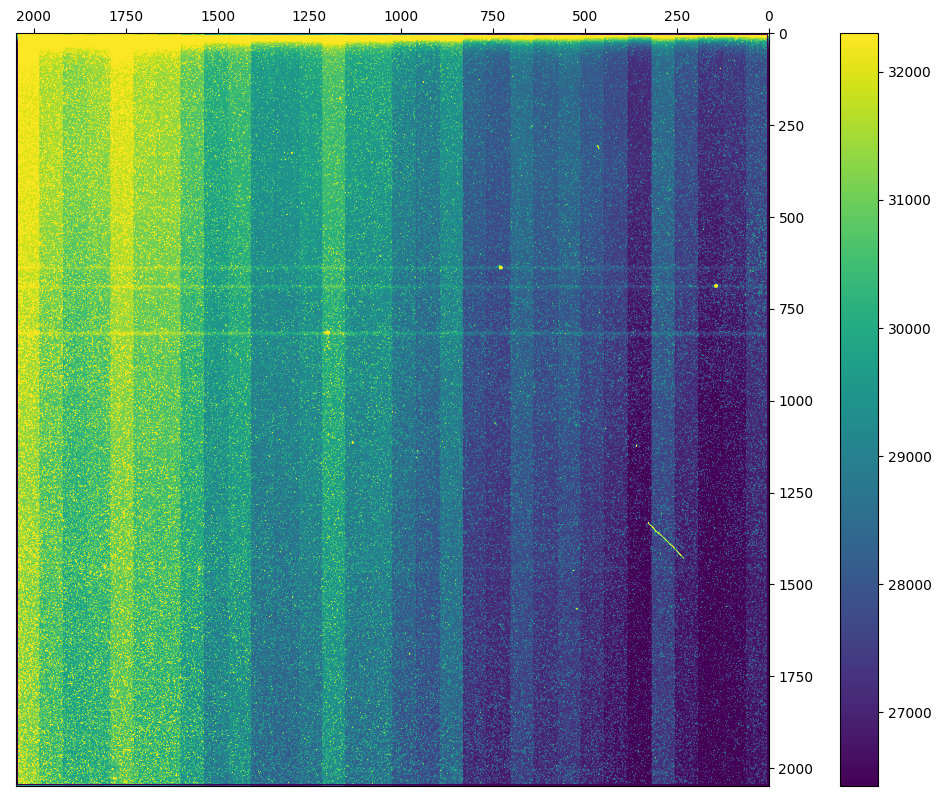}
    \end{center}
\end{minipage} \hfill
\begin{minipage}[c]{0.42\linewidth}
    \begin{center}
             \includegraphics[width=0.84\columnwidth]{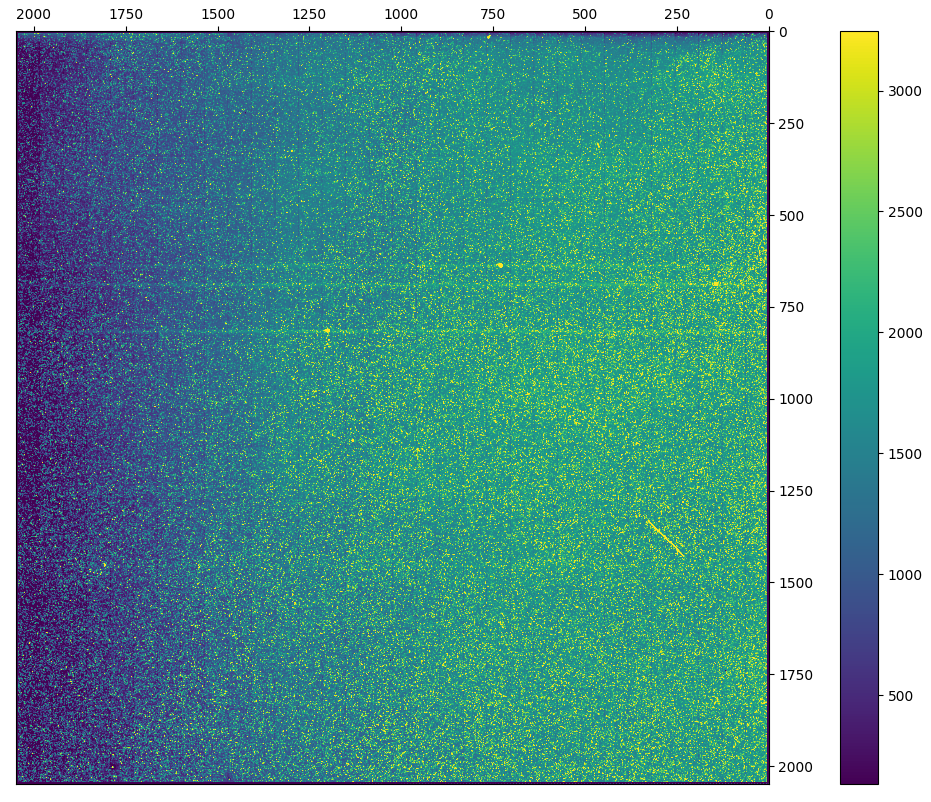}
    \end{center}
 \end{minipage} \hfill
\caption{Example of a frame before correction by reference pixels[left] and after correction by reference pixels [right]. The axes correspond to the pixel number on the detectors, and the color bars are in ADU. In this example, the bias has not been subtracted, hence, the main visible effect is an offset correction.}
\label{fig:corPixRef}
\end{figure}

The pixel gain is very sensitive to small perturbations of the bias voltages, induced by the environment. Some of these perturbations can be corrected when they affect several pixels in the same way. This is in particular the case for perturbations impacting the column preamplifiers, which affect an entire channel.
The noise generated at channel level is often called common mode noise and it can be reduced thanks to sets of reference pixels surrounding the matrix of active pixels. The reference pixels are identical to active pixels, except they are not connected to the sensitive layer. Since they have the same electronics as active pixels, they are subject to the same variations of the biases and temperatures as the active pixels. The aim is to correct the active pixels from these variations, by subtracting a linear combination of the signals measured by nearby reference pixels, as proposed by Bogna Kubik et al. \cite{Kubik2014} for H2RG sensors.
This correction is based on two steps. The first correction is applied to a whole readout channel, and the second one to each row individually. 
First, we select the top and bottom reference pixels of a readout channel, representing 256 pixels, and we compute the mean of their values. Then, we subtract this mean to all the pixels of the channel, including the reference pixels themselves.
Second, we select the left and right reference pixels of row L, plus 4 rows of reference pixels above and below row L, 9 rows in total, representing 54 pixels. We subtract their mean to row L. This second correction is computed with the left and right reference pixels previously corrected by the first correction. 
This method allows a reduction of the readout noise by $\sim 25 \%$ and is applied to each frame of the ramp. 
This correction also corrects the offset, channel by channel, if it was not subtracted beforehand. As we are computing our flux with differential ramps, constructed from the subtraction of two consecutive frames, this offset correction has no impact on our process. 
Figure \ref{fig:corPixRef} highlights the impact of the correction on the uniformity of a frame. However, the bias has not been subtracted beforehand here, hence, the main visible effect is an offset correction.

\subsubsection{Working with differential ramps}
\label{ssub:difframp}
As explained before, the ALFA sensor used in this study has a non-destructive readout. The signal continuously builds up in the pixels and it is read out every 1.3~s, making ramps whose slope is proportional to the flux received by the pixel.
Instead of fitting the ramps, which integrate the signal, we have chosen to fit the differential ramps constructed by subtracting the signal measured in frame k-1 to the signal measured in frame k. While the information contained in the differential ramps is equivalent to the one contained in the ramps, the differential ramps presents the advantage of being a series of independent measurements of about the same value, allowing a more direct statistical interpretation of the results. Fitting the differential ramps also requires one less parameter.

\begin{figure}[ht!]
\begin{minipage}[c]{0.46\linewidth}
    \begin{center}
             \includegraphics[width=\columnwidth]{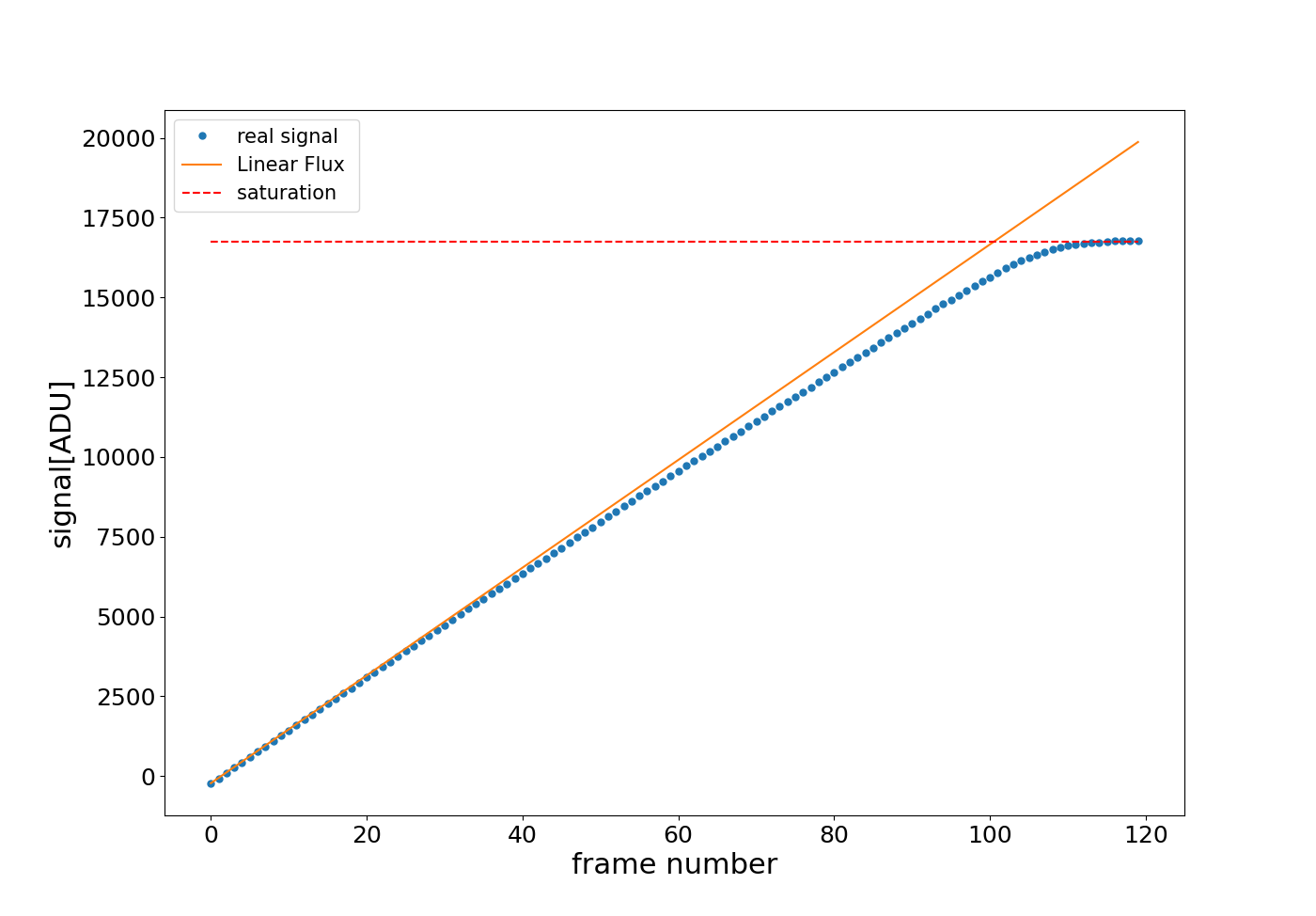}
    \end{center}
\end{minipage} \hfill
\begin{minipage}[c]{0.46\linewidth}
    \begin{center}
             \includegraphics[width=\columnwidth]{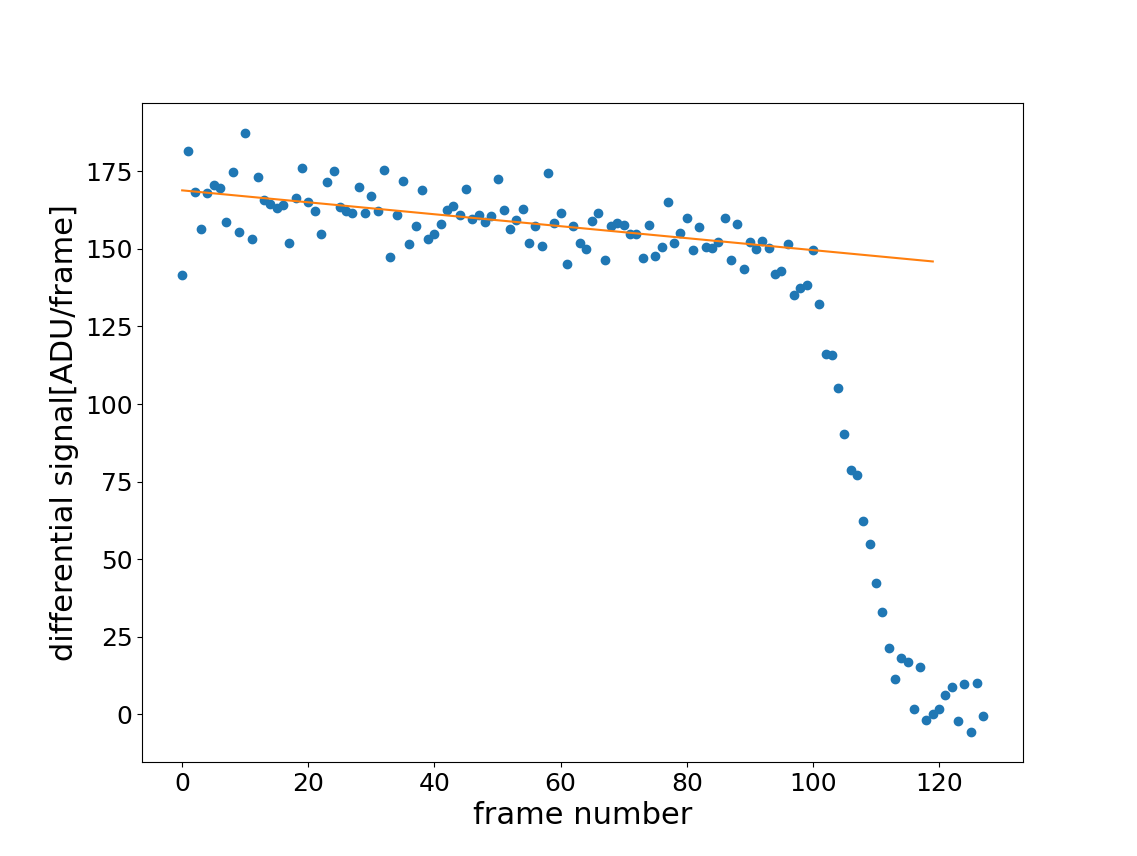}
    \end{center}
 \end{minipage} \hfill
\caption{Experimental ramp (blue point), linear extrapolation of the beginning of the ramp (orange) and saturation level (dashed red line) [left]. Differential ramp (blue points) and its fit (orange line) [right]}
\label{fig:rampdif}
\end{figure}

Figure \ref{fig:rampdif} gives an example of a ramp and the corresponding differential ramp. The left panel of figure \ref{fig:rampdif} shows a measured ramp (blue points) and the associated linear extrapolation of the beginning of the ramp (orange). The right panel shows the differential ramp constructed from the left curve.

\subsubsection{Identification and correction of pixels impacted by cosmics-rays}
\label{ssub:cosmic}

\begin{figure}[ht!]
\begin{minipage}[c]{0.46\linewidth}
    \begin{center}
             \includegraphics[width=\columnwidth]{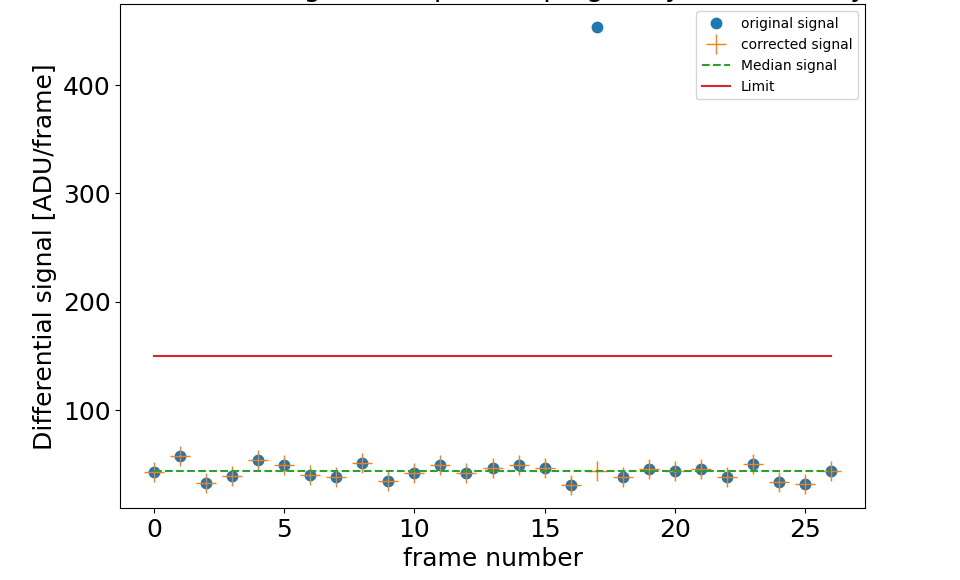}
    \end{center}
\end{minipage} \hfill
\begin{minipage}[c]{0.46\linewidth}
    \begin{center}
             \includegraphics[width=\columnwidth]{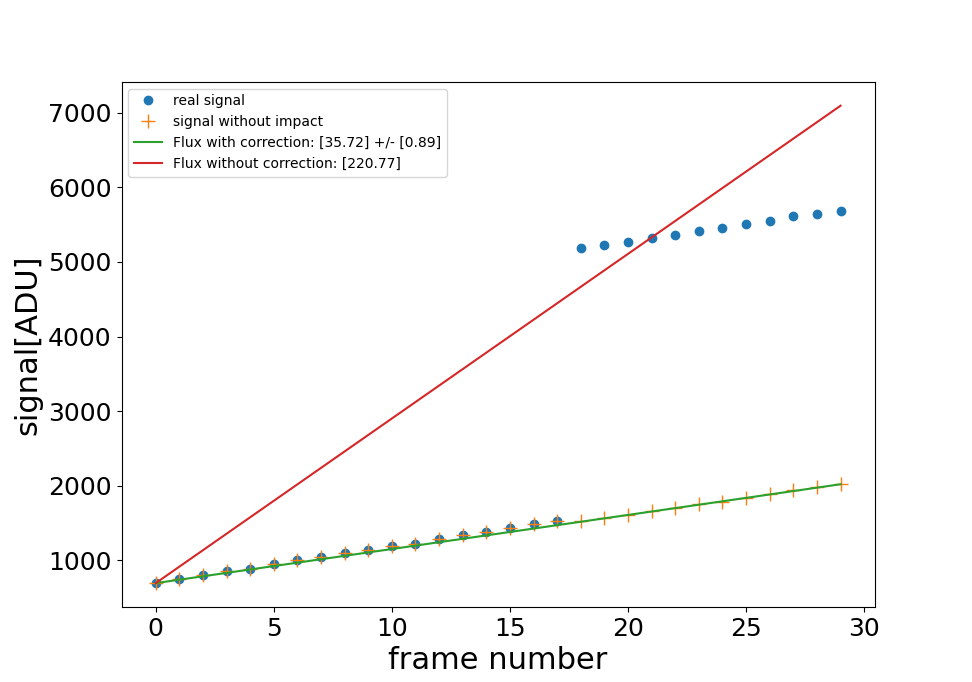}
    \end{center}
 \end{minipage} \hfill
\caption{Differential ramp of a pixel impacted by a cosmic-ray at frame number k=18, and limit to identify cosmic-rays. [left]. Ramp of a pixel impacted by a cosmic-ray at frame number k=18, and its fit before (orange) and after (green) correction [right].}
\label{fig:cosmic}
\end{figure}

Cosmic rays are high energy particles coming from outer space. They generate high-energy secondary particles in the atmosphere of the Earth, which can leave a track in the detector, especially for observatories located at high altitudes. Even if few cosmic-ray hits are detected on the calibration maps measured in the laboratory, a larger number of pixels will be impacted by comic-rays during observations. To avoid errors on the flux map, cosmic-rays hits have to be localized. We discuss here a method allowing to localize them and to recover the flux of pixels impacted by cosmic-rays. 
Cosmic-ray impacts have a particular signature on the ramps and on the differential ramps, by creating a big jump between the signal measured in two consecutive frames. This jump is responsible for a high variance on the differential ramp. However, erratic pixels, or pixels impacted by high fluxes can also show differential ramps with a high variance. To differentiate a pixel hit by a cosmic-ray from a pixel illuminated by a bright source, we rely on the ratio R of the variance over the mean of the differential ramp. 
For bright stars, this ratio should be small because of the large mean. On the contrary, this ratio will be large for pixels impacted by cosmic-rays and erratic pixels. Pixels with both the variance and R exceeding predefined limits (constructed from their respective histograms), have probably been impacted by cosmic-rays.
Once these pixels are selected, their differential ramps are tested individually. We compute the Median Absolute Deviation (MAD) of the differential ramp, and we define a limit to localize points very different from the median signal. This limit is plotted with a full row on the figure \ref{fig:cosmic} [left]. It is defined by:  
\begin{equation*}
    madLim = Signal_{median} + X \times MAD
\end{equation*}

Where "X" is a configuration parameter that will be defined with data acquired with the camera in its final configuration.  If we detect a point of the differential ramp with a value larger than madLim, it is probable that a cosmic-ray impinged it, hence we suppress this point from the differential ramp of the pixel, and thus avoid to consider it in the calculation of the flux. To differentiate them from erratic pixels, cosmic-rays impacts are identified only if a single point exceeds the limit madLim. If no point is detected above madLim, or several points are detected above madLim, the pixel is an erratic one, and it is rejected from the list of comic-rays hits.

Thanks to this method, we can detect cosmic-rays hits and recover the flux of the impacted pixels (Fig \ref{fig:cosmic}, right). The preprocessing pipeline returns the localization of the identified comic-rays hits, and the flux corrected from the impact of these cosmic-rays. 

\subsubsection{Flux measurement on the differential ramp}
\label{ssub:flux}
This is the most important work of the preprocessing, but also the longest if it is not optimized. It relies on the fit of differential ramps, and it is described in section \ref{sec:flux}.

\subsection{Products of the preprocessing pipeline}
\label{ssub:products}
The following maps are returned by the pipeline:
\begin{itemize}
    \item The position of saturated pixels.
    \item The position of pixels impacted by cosmic-rays.
    \item The number of frames used to fit the differential ramp of each pixel (section \ref{sub:linrange}).
    \item The signal received by each pixel as described in section \ref{sub:realramp}.
    \item The error on the signal as described in section \ref{sub:realramp}.
\end{itemize}

\subsection{Performances of the preprocessing pipeline}
\label{sub:perfo}

In terms of performances, we measured the time necessary to perform each step of the pipeline for the whole array, depending of the number of frames in the ramp, see table \ref{tab:time}. 
The pipeline has been written in Python, and is implemented in a laptop with a microprocessor Intel Core i7-10810U @ 1.10GHz.
The performances should thus be accelerated in the final process, thanks to the use of Julia language and a more powerful processor.

\begin{table}[ht!]
\begin{center}
\begin{tabular}{ | p{10cm} ||c| c| c |  c| } 
\hline
\textbf{Number of frames considered} & \textbf{7}  & \textbf{20} & \textbf{46 }& \textbf{90}\\
\hline
Exposure duration  [s] & 9  & 27 & 60 & 120\\
\hline
Finding saturated pixels [s]&  0.008& 0.008 & 0.008  & 0.02 \\
 \hline
Computing the limit of the fit [s]&0.3 &0.7 & 2 & 6  \\
\hline
Correction by reference pixels [s]& 0.8& 2.3 & 5& 9.7   \\ 
\hline
Creation of the differential ramp [s]& 0.2 &0.8  &2.0  & 3.6   \\
\hline
Computing output variables (Flux, Error, Cosmic-rays hits)[s]& 1.5 & 2.5 & 5.6& 19.4  \\
\hline
\textbf{Duration of the preprocessing} & \textbf{6} & \textbf{9}& \textbf{18} & \textbf{54} \\
\hline
\end{tabular}
\caption{Table of the characteristic processing time of the pipeline over the 4 millions of pixels}
\label{tab:time}
\end{center}
\end{table}

Table \ref{tab:time} demonstrates that, even with this configuration, the processing time is systematically shorter than the duration of the acquisition, which allows to finish the process before starting a new acquisition, as required. 

\section{Measuring the flux in near-real time}
\label{sec:flux}

While the flux received by a pixel is encoded in the slope of its ramp, many factors can perturb its measure.
First, the slope of the ramp decreases slowly due to the fact that the charges accumulated during an exposure tend to increase the effective capacitance of the pixel. Second, minute changes of the bias voltages produce slight variations of the gain, and consequently of the slope. Third, cosmic-rays may hit one or several pixels, producing a large number of charges and a jump in the ramp of these pixels. Finally, the sensor's non uniform illumination during sky exposures adds supplementary difficulties, like saturated pixels. Due to the wide field of view of CAGIRE, there will be many tens of saturated stars in standard exposures. It is nevertheless possible to get a correct estimate of the flux on these pixels, at the condition of measuring the slope of the ramp with only the frames recorded before saturation. It is thus necessary to evaluate, for each pixel, the range of frames to be used for the evaluation of the flux. 

We have discussed the management of saturation, gain changes and cosmic-ray hits in the previous section, so this section is specifically dedicated to the management of non-linearities in the evaluation of the flux received by a pixel.

\subsection{ Dealing with non-linearities with differential ramps}
\label{sub:diframp}

 The phenomenon of non-linearity can be seen in figure \ref{fig:rampdif}, where the signal on a pixel of an ALFA sensor shows a slight deviation from linearity (measured ramp in blue, and linear signal in orange), which has to be taken into account to estimate the flux from the ramp. Because of this deviation from linearity, the use of linear regression over the ramp is not adapted to measure the flux. Moreover, the statistical uncertainty on the points of the ramp increases with signal accumulation, because the detector noise is dominated by the photon Poisson noise. To overcome this problem, we decided to work with the differential ramps, constructed as the difference between two consecutive frames, as presented in the right panel of figure \ref{fig:rampdif}. 
  At some point (frame number 90 approximately on figure \ref{fig:rampdif}), the ramp reaches a plateau (red dashed line), corresponding to the saturation level. In the differential ramp, right panel, the saturation level corresponds to the sharp fall of the values. Close to this saturation level, the deviation from linearity is higher, leading us to select a usable range of the ramp (linear and non saturated) to evaluate the flux. 
The first step of our calibrations is the computation of the functional form of the non-linearity function and the applicable range of this function. These two quantities are calibrated thanks to calibration ramps under illumination provided by CEA-IRFU.

\subsection{Calibration of pixels non-linearity}
\label{sub:nonlin}

The choice of the fit of the differential ramps relies on the ability of the function to take the non-linearity into account. The fits are computed on differential ramps.
In order to choose, we compared three functions: 
 \begin{itemize}

\item A first order polynomial (a linear fit), where k is the frame number and the interesting parameter is  $a_0$, the signal in ADU/frame, which is proportional to the incident flux. $a_1$ is the non-linearity coefficient.
\begin{equation*}
    P_1(k) = a_0 + a_1 \times k
\end{equation*}
\item A second order polynomial (where the interesting parameter is $a_0$, the signal), 
\begin{equation*}
    P_2(k) = a_0 + a_1 \times k + a_2 \times k^2
\end{equation*}
\item A non-polynomial function using a coefficient of non-linearity, defined by:  
\begin{equation*}
    F_{\delta}(k) = \frac{a_0}{\left(1 - \delta \times a_0 \times k\right)^2}
\end{equation*}
where $a_0$ is the signal we are looking for, and $\delta$ the non-linearity coefficient.

 \end{itemize}
As the goal of the preprocessing is to be precise on the estimation of the signal but also fast and robust, the linear fit would be preferred. However, functions with more parameters could have the advantage of providing more accurate fits. 
To compare the efficiency of these functions to fit our data, we compared the medians of several quantities: the signal estimated by fitting a differential ramp, the variance on the fitted signal, the $\chi^2$ associated to the fits and the normalized $\chi^2$ associated to the fits. The normalized $\chi^2$ is computed by dividing the $\chi^2$ value by the number of degrees of freedom of the function (number of points of the ramp fitted minus the number of parameters of the fit, minus one). The results are computed on all the active pixels of the detector, and the medians of these values over the detector are summarized in table \ref{tab:chi}.
The three functions have been used to fit the differential ramps over 80\% of the dynamic range of the pixels.

\begin{table}[ht!]
    \centering
    \begin{tabular}{|l|p{2.5cm}|p{2.5cm}|p{2.5cm}|}
    \hline
     Function & $2^{nd}$ order pol. & $1^{st}$ order pol. & $F_{\delta} $ \\
    \hline
     Number of fitted parameters & 3 & 2 & 2 \\
    \hline
    Signal [ADU/frame] (median value) & 193.5 & 193.3  &  193.5  \\
    \hline
     Variance [ADU²/frame²] (median value) &  7.4&  3.1 & 3.4  \\
     \hline
    $\chi ^2$ (median value) & 62.6 &  63.3 &  63.3 \\
    \hline
    Normalized $\chi ^2$ (median value) & 0.947 &  0.942 &  0.927 \\
    \hline
    
    \end{tabular}
    \caption{Comparison of the values of signal, variance, $\chi ^2$  and normalized $\chi ^2$ for the three non-linearity functions studied here. The values are the median values over all the active pixels of the detector.}
    \label{tab:chi}
\end{table}

Table \ref{tab:chi} highlights values of the signal and of the $\chi^2$ that are very similar for the three functions, providing no strong argument in favor of one function or another, from their sole capacity to fit the data. Looking at Table \ref{tab:chi}, we see that the median $\chi ^2$ of the three functions,  are very close, meaning that the $2^{nd}$ order polynomial and $F_{\delta}$ do not provide significant improvement, compared to the linear fit, which is the simplest function. We thus decided to fit the differential ramps with a linear fit to compute the signal. 


The $a_1$ coefficient of the linear fit has units of $\frac{ADU}{frame^2}$. It is the consequence of the increase of the effective capacitance of the pixel while accumulating charges, which explains the deviation from linearity. We measured a median value of this coefficient of -0.24 ADU/frame$^{2}$ for active pixels. The negative sign reflects the capacitance increase, and thus the decrease of the number of ADU encoded for the same number of charges received. The role of this parameter in our process is described in section \ref{sub:linrange}.

\subsection{ The fitting range of the ramps and the non-linearity coefficient}  
\label{sub:linrange}

To determine the fitting range of the ramps, we worked with ramps provided by CEA-IRFU. These ramps have been acquired with the detector under constant illumination during more than 4 min, to let the detector reach saturation.
We fit the differential ramps with a first order polynomial, and compute the signal as the zero$^{th}$ order parameter of the fit, called $a_0$. The first order parameter is called $a_1$ and is representative of the loss of signal between two consecutive frames caused by the non-linearity of the ramp. We then looked at the variations of the estimated signal with the range of the ramp fitted, taken as a fraction of the saturation level. The results are presented in Table \ref{tab:flux} and show stable signal $a_0$ estimates up to 80-90\% of the saturation.

\begin{table}[ht!]
    \centering
    \begin{tabular}{|c||c|c|c|c|}
    \hline
    Saturation level fraction [\%] & signal [ADU/f] & Error [ADU/f] & Dev. from lin [\%] &Number of frames \\
    \hline
    10 \%& 193.255 & 4.9 & 0.3  &10\\
    \hline
    20 \%  & 193.164&  3.5& 0.9 &18 \\
    \hline
    30 \% & 193.135& 3.0& 1.5 & 27\\
    \hline
    40\% & 193.129& 2.6 & 2.1 &35\\
    \hline
    50 \%& 193.244&  2.4& 2.8  &44\\
    \hline
    60\%& 193.358& 2.4& 3.4 & 53\\
    \hline
    70 \%& 193.350& 2.3 & 4.0 &62\\
    \hline
    80 \%& 193.234& 2.4 & 4.7 &71 \\
    \hline
    90 \%& 193.214&2.6 & 5.3  &80\\
    \hline   
    \end{tabular}
    \caption{Signal estimation as a function of the maximum level of the ramp fitted, the associated error on the signal coefficient and the deviation from linearity at this level. The last column represents the number of frames used to compute the fit.}
    \label{tab:flux}
\end{table}

\begin{figure}[!ht]
     \begin{center}
             \includegraphics[width=11cm]{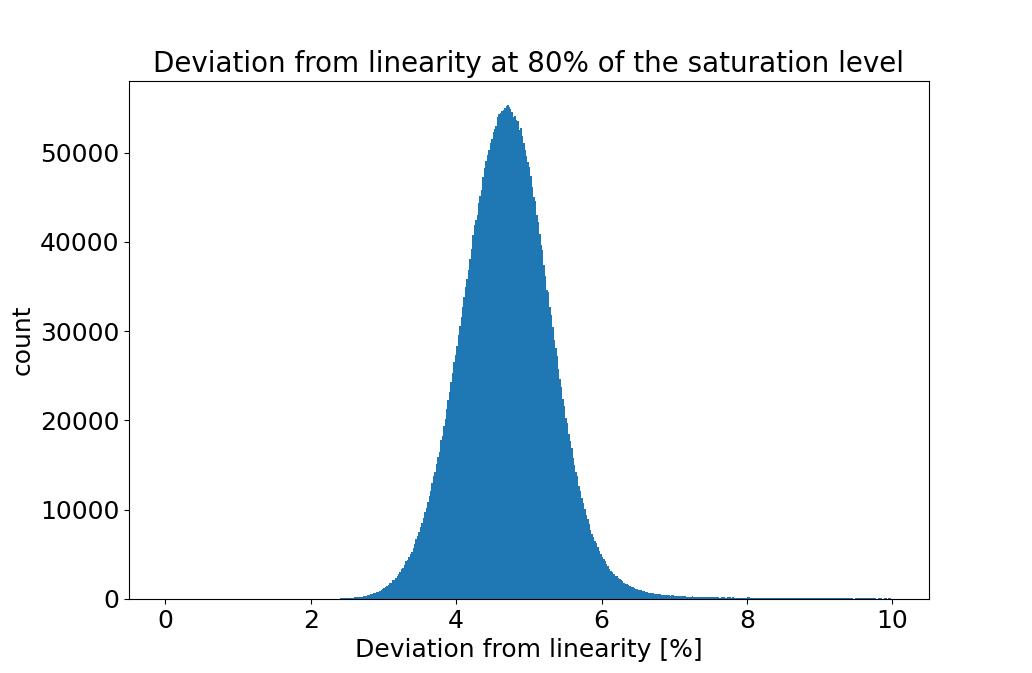}
         \end{center}
 \caption{Histogram of the absolute value of the deviation from linearity at 80\% of the saturation level}
\label{fig:devlin80}
\end{figure}

Table \ref{tab:flux}, highlights very stable signal coefficients with a fit by a first order polynomial, with variations smaller than 0.1\%.
These results confirm our choice of using a first order polynomial to fit the differential ramps. Moreover, we see that below 80\% of the dynamic range, the deviation from linearity is smaller than 5\% for most pixels, which is not the case at 90\% of the saturation. Figure \ref{fig:devlin80} shows the histogram of deviation from linearity for all active pixels at 80\% of the dynamic range. It shows that pixels can have different behaviours. Because of this dispersion better results are obtained when the fitting range is defined as a fraction of the saturation level, instead of a deviation from linearity. If we choose a fitting range based on a given deviation from linearity (e.g. 5\%), for some very linear pixels the fitting range could exceed the saturation level, leading to an erroneous value of the signal. 
We thus decided to fit the ramp until 80\% of the saturation level to be sure to be in the close-to-linear domain of the ramps.

Moreover, the fit of differential calibration ramps under illumination with a first order polynomial function over 80\% of the saturation range, allows us to construct a map of the non-linearity coefficient for each pixel of the detector. However, the coefficients $a_0$ and $a_1$ are not ``universal'' as they depend on the illumination. Specifically
$a_0$ depends on the signal, while $a_1$ depends on the signal accumulated by the pixel. 
A way to get rid of this dependence is to construct the parameter $\gamma = \frac{a_1}{a_0^2}$, which can be shown to encode the non-linearity intrinsic to each pixel (see Annex \ref{ann:gamma}). We thus use calibration ramps to construct a map of $\gamma$ (figure \ref{fig:gamma} of annex \ref{ann:gamma}).
In the next section we discuss how this information is used to process ramps measured on the sky.

\subsection{Measuring the signal with real ramps}
\label{sub:realramp}

The first step is to identify for each pixel, the frame number at which it reaches 80\% of its saturation level, using calibration maps described in section \ref{sub:sat}. This limit, labelled k$_{\rm 80}$ is different for each pixel.
The second step is to compute the differential ramp and to estimate the signal, using a linear fit of the differential ramp.

\begin{equation}
    d_k = a_0 + a_1 \times k
\label{eq:dk}
\end{equation}

For the stability of the signal estimate, we want to estimate only the signal $a_0$, considering the non-linearity as a known parameter.
Relying on the non-linearity parameter $\gamma = \frac{a_1}{a_0^2}$, computed during the calibrations, equation \ref{eq:dk} can be rewritten as follows. 

\begin{equation*}
    d_k = a_0 + a_0^2 \times \gamma \times k
\end{equation*}

where $k$ is the number of the frame considered, $a_0$ is the signal to be estimated, and  $\gamma$, is the coefficient of non-linearity described in the previous section.  After some computation we find that the signal on sky data can be computed by solving the equation: 

\begin{equation*}
    A \times a_0^2 + B \times a_0 +C = 0
\end{equation*}
where A = $\gamma \times \frac{ (N+1)}{2}$ ; B=1  and C = - $\frac{1}{N} {\sum_{k=1}^{N} d_k}$. N is the number of frames of the ramp used in the calculation. In the end, we obtain the signal $a_0$ in ADU/frame :
\begin{equation}
    a_0 = \frac{-B + \sqrt{B^2 - 4AC}}{2A}
\label{eq:flux}
\end{equation}
The error on this coefficient is defined by : 
\begin{equation*}
    Error(a_0) = \sqrt{\frac{ variance(d_k)}{N \times \left(B^2 - 4AC\right)}}
\end{equation*}

The computation of $a_0$ and its error are performed on the differential ramps up to k$_{\rm 80}$, for each pixel independently. Figure \ref{fig:rampdif}, shows an example of a differential ramp (blue points) fitted by this method (orange line). In this figure k$_{\rm 80}$ = 81. More results on sky images are given in section \ref{sec:sky}.

\section{Calibration maps}
\label{sec:badpix}

Calibration maps are created beforehand, in order to localize bad pixels (dead, erratic...), to measure the saturation level, to construct the bias map and to evaluate the usable -- sufficiently linear and unsaturated -- range of  the ramp for each pixel. These calibration maps, which are necessary for the preprocessing pipeline, will be acquired at the ``Centre de Physique des Particules de Marseille'' (CPPM) during the calibrations of the CAGIRE sensor.

\subsection{The saturation level
\label{sub:sat}}
First, a map of the saturation level per pixel is needed to evaluate the range of the ramp on which each pixel can be operated. It is also a useful map to localize the pixels saturated during an acquisition. This map is obtained by illuminating continuously the detector during the acquisition of a single ramp. The detector has to be illuminated until saturation to compute the saturation level of each pixel. An example of a saturation map acquired with the calibration data provided by CEA-IRFU is given figure \ref{fig:calib}.

\subsection{Bias map}
\label{sub:bias}

\begin{figure}[ht!]
\begin{minipage}[c]{0.43\linewidth}
    \begin{center}
             \includegraphics[width=\columnwidth]{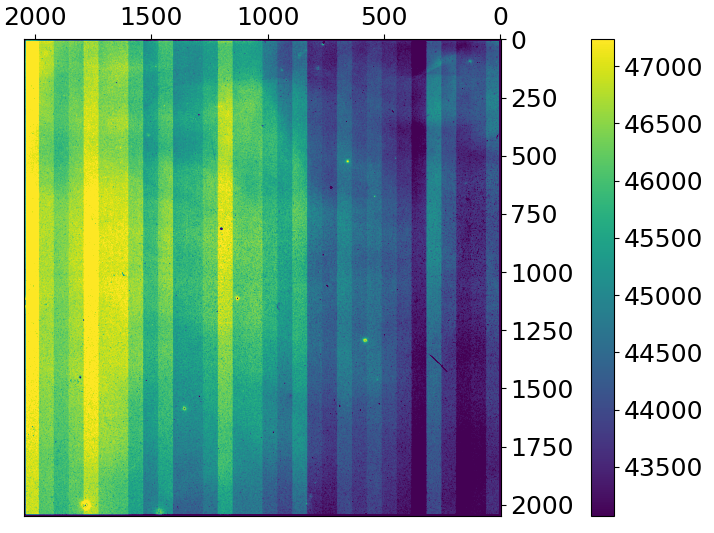}
    \end{center}
\end{minipage} \hfill
\begin{minipage}[c]{0.43\linewidth}
    \begin{center}
             \includegraphics[width=\columnwidth]{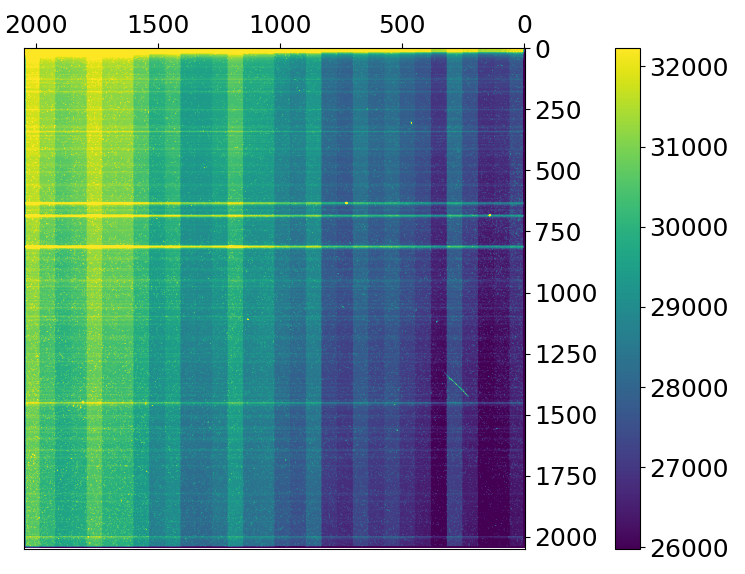}
    \end{center}
 \end{minipage} \hfill
\caption{Example of a saturation map. The saturation level is expressed in ADU. The bias has not been removed. [left]. Example of a bias map. [right]}
\label{fig:calib}
\end{figure}

Before correcting the frames from the common mode noise, we need to subtract a Master Bias, measured during calibration tests. A pseudo-bias is computed as the median signal of few tens of dark acquisitions of one frame. This bias is required to deal with the first frame of a ramp. An example of a pseudo-bias map computed from 2 dark files  provided by CEA-IRFU is given figure \ref{fig:calib} [right].

\subsection{Non-operational pixels}
\label{sub:nonoppix}

We also need a map of the non-operational pixels to avoid misidentifications. Pixels can be considered as non-operational if they are dead or erratic. Hot pixels will be unusable only if they saturate in the very first frames of the ramp. Cold pixels are problematic if their limited response to light introduces a signal very different from their neighbours. Knowing their position can be useful for the astronomy pipeline. We explain in the next paragraphs how we localize and characterize these pixels thanks to calibration maps in the dark and under illumination. In the end, the map returned to the astronomy pipeline consists of a single map where all non-operable pixels are flagged in the same way.

\begin{figure}[ht!]
\begin{minipage}[c]{0.47\linewidth}
     \begin{center}             
            \includegraphics[width=\columnwidth]{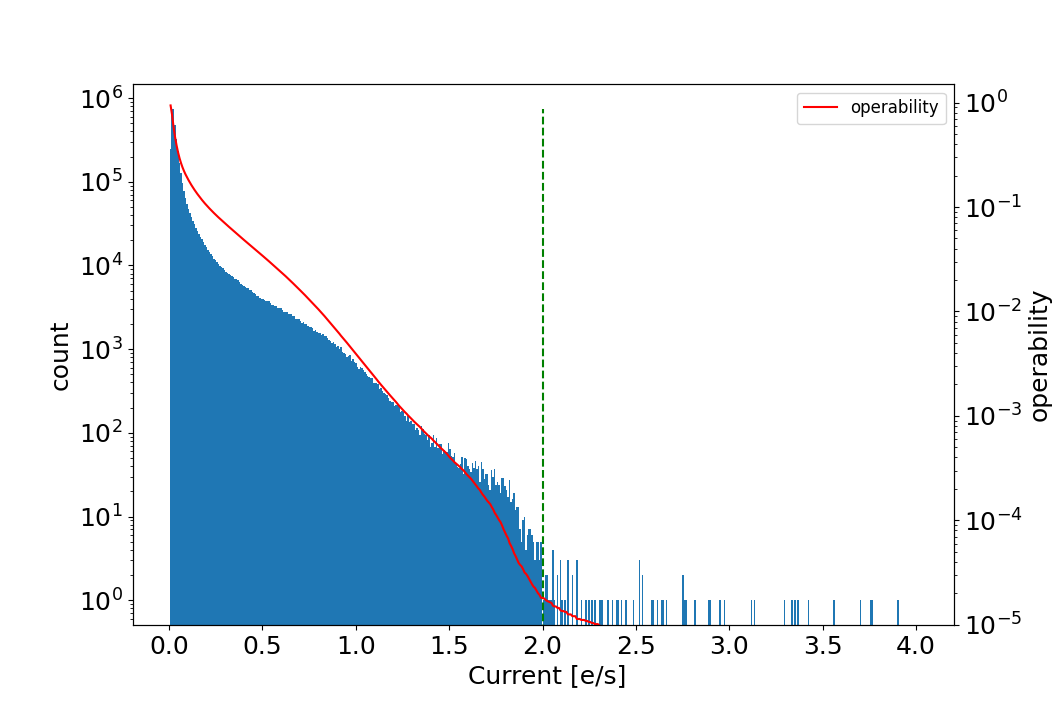}
         \end{center}
   \end{minipage} \hfill
\begin{minipage}[c]{0.45\linewidth}
     \begin{center}
             \includegraphics[width=\columnwidth]{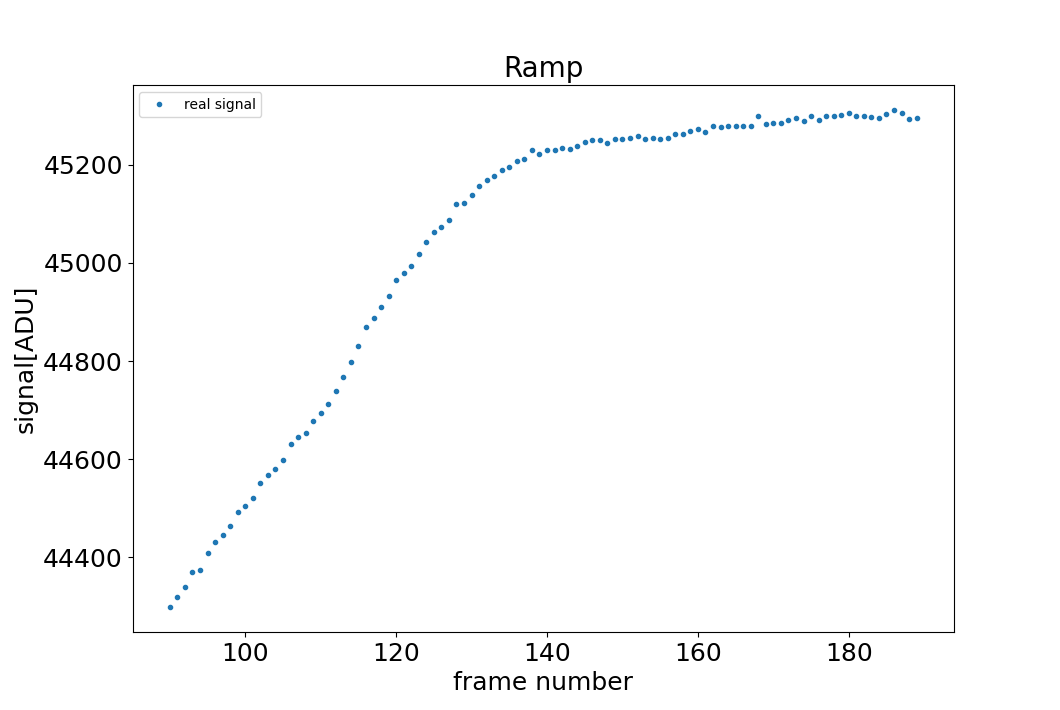}
         \end{center}
   \end{minipage} \hfill
 \caption{Left: histogram of the dark current for a 60266 second long exposure and limit of hot pixels (green dashed line). The red plot represents the "operability" (the fraction of pixels with a dark current larger than the abscissa), which can be read on the right vertical scale. Right: ramp of a hot pixel. }
\label{fig:hot}
\end{figure}

\textbf{Hot pixels} are pixels with an abnormally high signal in the dark. We notice that most of them are usable while being illuminated. To find hot pixels, we compute the dark current, and we study pixels revealing a very high dark current, or saturating pixels. Only saturating pixels and pixels with a dark current larger than to 2\,electrons/s will be considered as unusable (see fig \ref{fig:hot}). These pixels represent less than 0.001\% of the active pixels. We also plotted here an "operability" curve on the histogram, representing the fraction of events larger than the value of x-axis. This curve highlights that very few pixels overcome the limit of 2 e-/s.

\begin{figure}[ht!]
\begin{minipage}[c]{0.48\linewidth}
     \begin{center}             
            \includegraphics[width=\columnwidth]{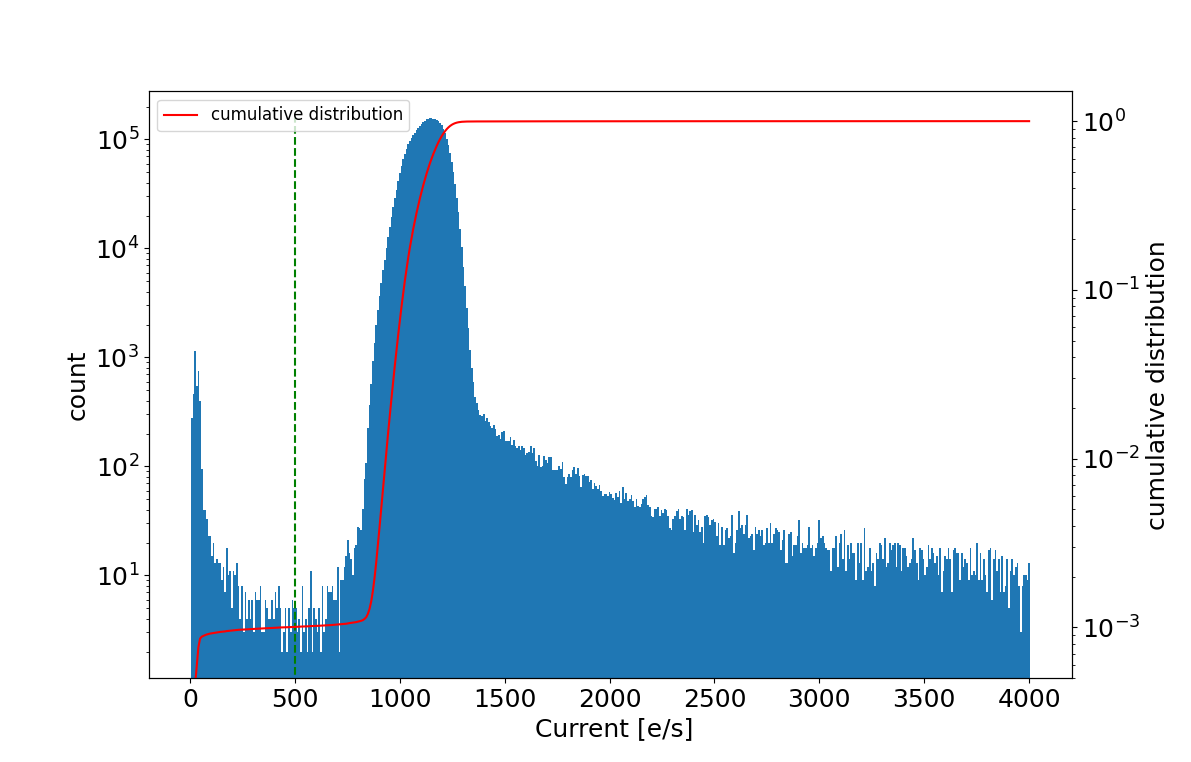}
         \end{center}
   \end{minipage} \hfill
\begin{minipage}[c]{0.48\linewidth}
     \begin{center}
             \includegraphics[width=\columnwidth]{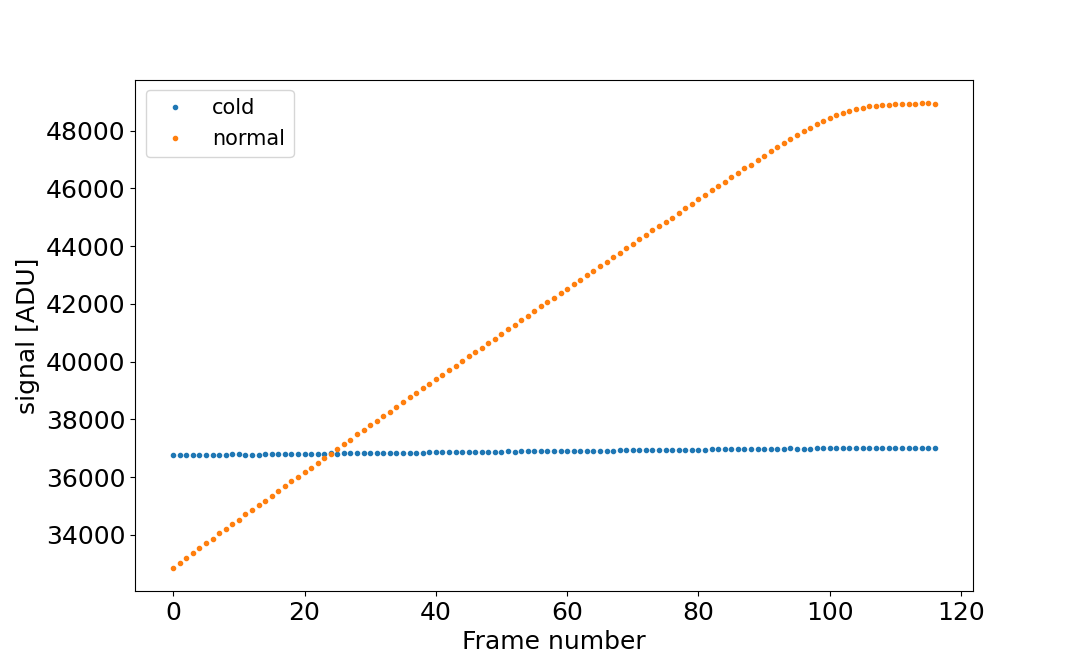}
         \end{center}
   \end{minipage} \hfill
 \caption{Left: histogram of the signal under illumination and the limit of dead and cold pixels (green dashed line). The red plot represents the cumulative distribution (because the pixels of interest are below the histogram maximum), which can be read on the right vertical scale. Right: ramp of a cold pixel (blue ramp) and ramp of an operable pixel (orange ramp).}
\label{fig:cold}
\end{figure}

\textbf{Cold pixels and dead pixels} on the contrary, are pixels with an abnormally low response to light. To find them, we compute the signal received by the detector while being illuminated by a constant flux, and we plot the histogram of this signal. Pixels with a signal smaller than 500\,electrons/s (see fig \ref{fig:cold}), are considered as cold or dead. These pixels represent 0.11\% of the active pixels. We also plotted here the cumulative distribution of the histogram, representing the fraction of events smaller than the x-axis value. This curve highlights that very few pixels are below the limit.

\begin{figure}[ht!]
\begin{minipage}[c]{0.48\linewidth}
     \begin{center}             
            \includegraphics[width=\columnwidth]{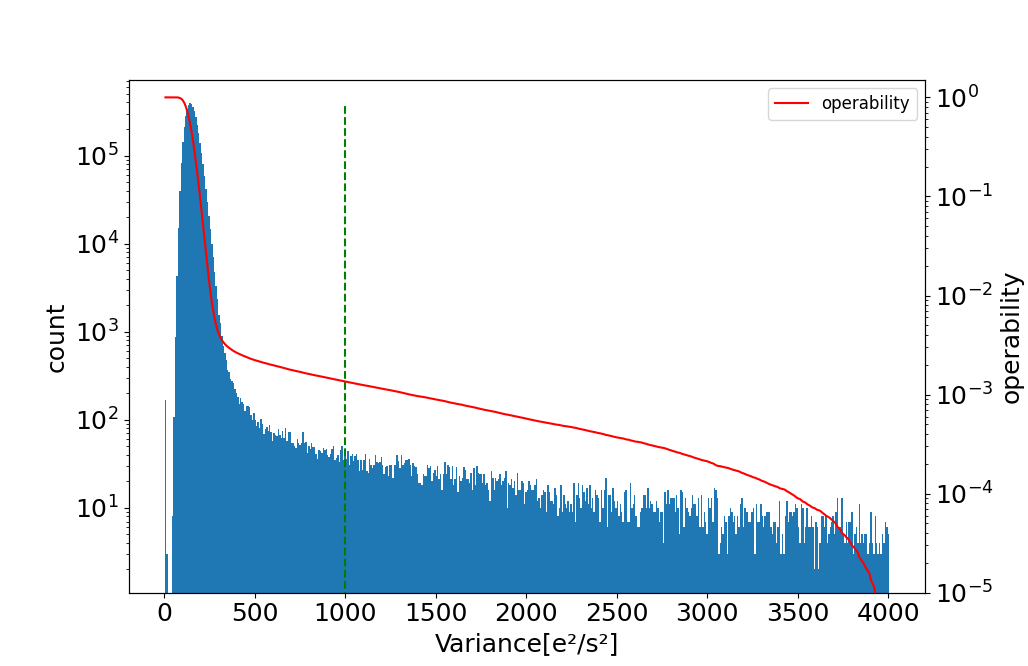}
         \end{center}
   \end{minipage} \hfill
\begin{minipage}[c]{0.47\linewidth}
     \begin{center}
             \includegraphics[width=\columnwidth]{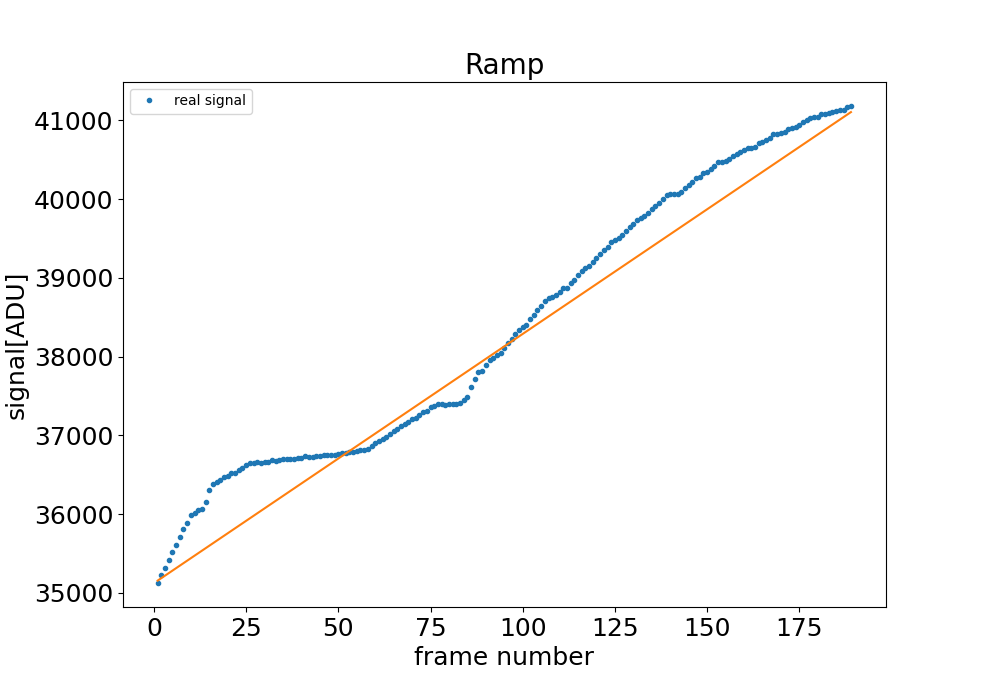}
         \end{center}
   \end{minipage} \hfill
 \caption{Left: Histogram of the variance on signal and limit of erratic pixels (green dashed line). The red plot represent the operability, which can be read on the right vertical scale. Right: ramp of an erratic pixel }
\label{fig:erratic}
\end{figure}

\textbf{Erratic pixels} are pixels with an unusual response to light, very dispersed, which can have different behaviours. To find them, we compute the variance of the signal, for each pixel. Erratic pixels are pixels with a variance larger than 1000 electrons²/s²(fig \ref{fig:erratic}). This limit is a preliminary result to define maps to test our preprocessing pipeline. A study of these pixels with a response very different from a line is currently conducted to better understand and localize them. With this limit of 1000 electrons²/s², erratic pixels represent 0.25\% of the active pixels. We also plotted here the "operability" curve on the histograms, representing the fraction of events larger than the value on the x-axis. This curve highlights the fact that very few pixels have an erratic response, according to our definition.

In the end, very few pixels are considered as non-operational. They represent $\simeq $ 0.35\% of the active pixels.
Using them could produce false detections so that their localization has to be provided to the astronomy pipeline, to avoid any misidentification based on the signal map.


\section{Application to sky images}
\label{sec:sky}

In order to validate the preprocessing pipeline, we tested it on images of the sky acquired with RATIR. These tests allow to confirm that the maps returned by the process are usable by the astronomy pipeline, but also to ensure that the signal returned on known stars is in line with the one listed in catalogs like the \textit{2MASS} All-Sky Catalog of Point Sources \cite{Cutri2003}. It is also a way to test that the preprocessing works properly on ramps that are dominated by the sky background, inhomogeneously illuminated and contaminated by cosmic-rays.

\subsection{Pre-processing of RATIR images} \label{sub:ratir}
The preprocessing pipeline described here has been tested on images of the sky taken with RATIR (the Reionization And Transients InfraRed camera). RATIR is a multi-channel optical and near infrared imager operating at the robotic 1.5-meter Johnson telescope at the Observatorio Astronómico Nacional in Mexico. This imager is able to acquire images in the r,i,Z,Y,J and H bands simultaneously. It is composed of two H2RG detectors, of 2040 by 2040 active pixels, surrounded by 4 rows and lines of reference pixels\cite{Butler2012}. For the sake of testing CAGIRE's preprocessing, RATIR data have been acquired in up-the-ramp mode.

\begin{figure}[ht!]
\begin{minipage}[c]{0.50\linewidth}
    \begin{center}
             \includegraphics[width=\columnwidth]{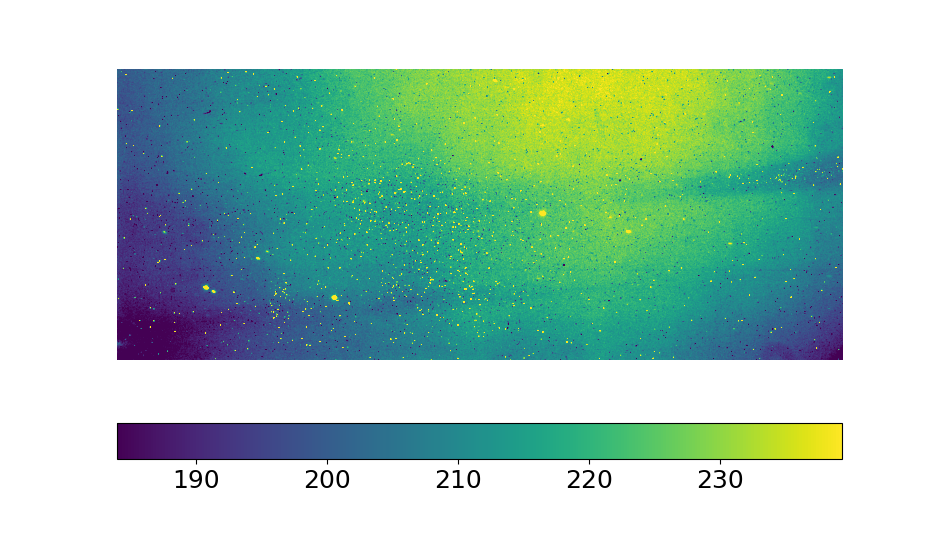}
    \end{center}
\end{minipage} \hfill
\begin{minipage}[c]{0.50\linewidth}
    \begin{center}
             \includegraphics[width=\columnwidth]{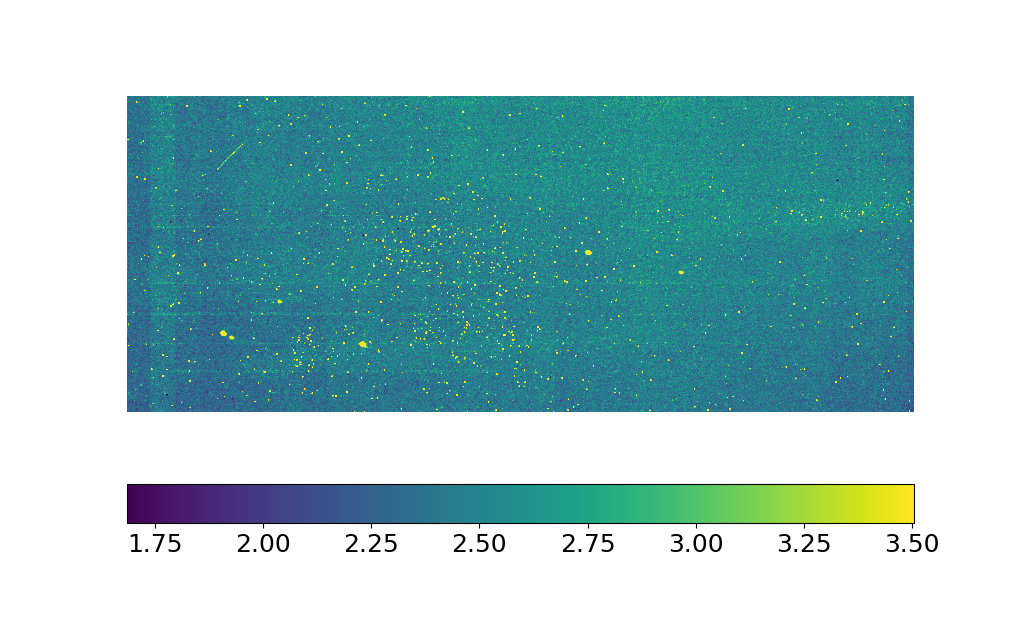}
    \end{center}
 \end{minipage} \hfill
\caption{Example of a flux map and the associated error in the H channel}
\label{fig:flux}
\end{figure}

The mode of operation with RATIR is thus very similar to CAGIRE, and it is located at the same site. The images acquired with RATIR are divided into two zones, one for the H photometric band and one for the J photometric band. We process here the whole image but distinguish stars observed in the J channel and stars observed in the H channel. Figure \ref{fig:flux} represents the signal and the associated error computed for the H channel.

The data under study are composed of ramps of 30 frames, acquired every 1.45\,s. We used these data to compute the signal of ramps taken with RATIR, using CAGIRE's preprocessing pipeline. We have to notice here that we could not apply the steps of the preprocessing that use calibration maps. It means that we could not compute the level at 80\% of the saturation and all pixels are processed in the same way. Moreover pixels saturating or non-operable pixels could not be removed. However we could apply the correction by reference pixels and construct all the output maps except the map of saturated pixels. Despite this lack of calibration maps, the process  of ramps acquired with RATIR has not raised any particular issue. We are thus confident of the robustness of the pre-processing. The following section is not part of the pre-processing pipeline but aims to test the relevance of our outputs.

\subsection{Results} \label{sub:ratirresults}
After the pre-processing, we used the map of the signal, figure \ref{fig:flux}, to extract the sources with {\fontfamily{pcr}\selectfont Source Extractor} \cite{Bertin1996}, and we cross-matched them with the catalog \textit{2MASS} All-Sky Catalog of Point Sources,\cite{Cutri2003} thanks to \textit{Astrometry.net} \cite{Lang2010}.

Then, we compared the number of sources identified with the number of source detected, and the magnitude measured by {\fontfamily{pcr}\selectfont Source Extractor} with the magnitude measured in the catalog \textit{2MASS}. The magnitudes estimated by {\fontfamily{pcr}\selectfont Source Extractor} have been computed without reference to the zero magnitude of the instrument. To make them comparable, we added an offset corresponding to the magnitude difference between \textit{2MASS} and {\fontfamily{pcr}\selectfont Source Extractor} for the brightest star detected. 

\begin{table}[ht!]
    \centering
    \begin{tabular}{|p{8cm}|p{3cm}|}
      \hline
      Number of extraction by {\fontfamily{pcr}\selectfont Source Extractor} & 31  \\
      \hline 
      Number of extraction associated to \textit{2MASS} catalog   & 28   \\
      \hline
      percentage of good extraction [\%] & 90.3  \\
      \hline
     \end{tabular}
    \caption{Number of sources extracted in our map, number of extractions associated to \textit{2MASS} catalog and corresponding percentage of good extractions. }
    \label{tab:results mean}
\end{table}

We notice here 90.3\% of good detection. However, the bad detections are located on a corner of the detector presenting a ghost image, and could probably be avoided using the right calibration files. 

  \begin{figure}[ht!]
        \centering
        \includegraphics[width=17cm]{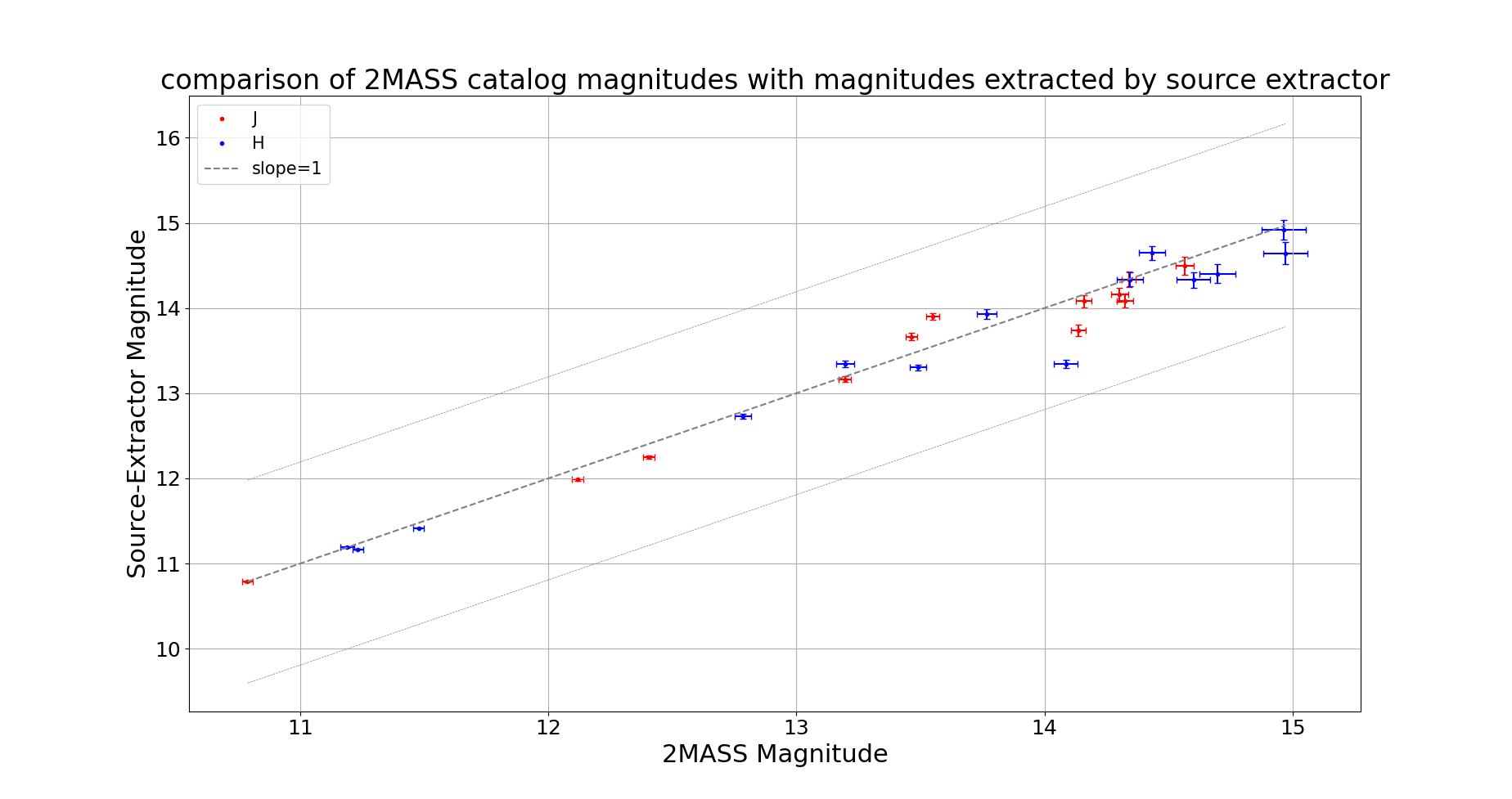}
        \caption{Comparison of the magnitudes measured by source extractor with magnitudes in the \textit{2MASS} All-Sky Catalog of Point Sources}
        \label{fig:2mass}
    \end{figure}

  The result plotted Figure \ref{fig:2mass} depicts the magnitude measured by {\fontfamily{pcr}\selectfont Source Extractor} versus the magnitude of \textit{2MASS} catalog. Red and blue points are for J and H bands respectively. The magnitudes estimated are in line with \textit{2MASS} results. They are not intended to be accurate magnitudes, however, because they are estimated with a single image, without correction of the sky background and with our very limited experience of {\fontfamily{pcr}\selectfont Source Extractor}. The astronomy pipeline of CAGIRE will indeed provide much more accurate magnitudes.
 Nevertheless, these first results show that the preprocessing fulfills its role, by providing images cleaned from detector's features and ready for the astronomical analysis, starting from realistic raw sky images.
 Finally, CAGIRE is composed of pixels larger than RATIR's, it has a larger field of view and uses a different sensor. The preprocessing will thus need to be adapted to the sky images acquired by CAGIRE.

\section{Conclusion and perspectives}
\label{sec:conclusion}

We have discussed here a method to recover the sky signal, based on the fit of differential ramps, and its integration into the preprocessing pipeline of CAGIRE. The pipeline has to be fast enough to process a ramp before the acquisition of a new one.
We have shown here that it fulfills these requirements, returning maps usable by the astronomy pipeline, in a context of ramps dominated by the sky background.
The input maps of the pipeline (saturation, bias and non-operable pixels) and their acquisition have been described. We have also presented the correction of the common mode noise thanks to a combination of the signal measured by the reference pixels. Our method allows to recover the signal of pixels impacted by cosmic-rays, thanks to the computation of the MAD of the differential ramp.  A study of the speed performances of the pipeline has been conducted to attest the compliance with the requirement for a fast processing. The data provided by this process,  were finally tested on sky images. We have shown that we are able to detect the objects with few errors and to compute magnitudes in line with the ones from \textit{2MASS} catalog. For the future, we still need to study the effects of persistence on a short time scale ($\sim 1$ minute, comparable to the duration of a ramp), particularly for saturating pixels. To do so, we will base our work on the studies of T. Le Goff \cite{Legoff2020}. The impact of persistence will nevertheless be mitigated thanks to the use of a proper dithering strategy.  
Finally, our analyses, made with files acquired at CEA-IRFU, confirm that the ALFA detector fulfills CAGIRE requirements, with a good linearity and homogeneity, and  very  few defective pixels.

\section*{Acknowledgments}

The calibration data used in the first parts of this paper were acquired at CEA by O. Boulade. The data has been provided to IRAP in agreement with ESA. CAGIRE is partly funded by the French Centre National d'Etudes Spatiales (CNES). The PhD contract of A. Nouvel de la Flèche is financed by CNES and LYNRED. This work has been partially supported by the LabEx FOCUS ANR-11-LABX-0013.

The on sky data used in this paper were acquired with the RATIR instrument, funded by the University of California and NASA Goddard Space Flight Center, and the 1.5-meter Harold L. Johnson telescope at the Observatorio Astronómico Nacional on the Sierra de San Pedro Mártir, operated and maintained by the Observatorio Astronómico Nacional and the Instituto de Astronomía of the Universidad Nacional Autónoma de México. Operations are partially funded by the Universidad Nacional Autónoma de México (DGAPA/PAPIIT IG100414, IT102715, AG100317, IN109418, IG100820, and IN105921). We acknowledge the contribution of Leonid Georgiev and Neil Gehrels to the development of RATIR. 

\bibliographystyle{spiebib} 
\bibliography{b} 

\clearpage
\appendix
\section{Stability of the non-linearity coefficient \texorpdfstring{$\gamma$ }{a}}
\label{ann:gamma}

We consider here a differential ramp characterized by the equation: 

\begin{equation*}
    d_k = A_0 + A_1 \times k
\end{equation*}

We can define the signal value as: 

\begin{equation*}
   S_k = \sum_{j=1}^{k} d_j = A_0 \times k + A_1 \times \sum_{j=1}^{k} d_j = A_0 \times k + A_1 \times \frac{k(k+1)}{2}
\end{equation*}

The non-linearity is defined by the relative deviation of $S_k$ from the theoretical value, $A_0 \times k$ : 

\begin{equation*}
   Non-Linearity(k) = \frac{S_k - A_0 \times k}{A_0 \times k} = \frac{A_1}{A_0} \times \frac{(k+1)}{2}
\end{equation*}

If we do the same computation with a different signal $A'_0 = x \times A_0 $, we have :

\begin{equation*}
   Non-Linearity'(k) =  \frac{A'_1}{A'_0} \times \frac{(k'+1)}{2}
\end{equation*}

Considering that, far from saturation level, the non-linearity is proportional to the accumulated signal $S_k = A_0\times k $,  we have $Non-Linearity = Non-Linearity' $ at the same accumulated signal level.  We also have $A'_0 = x \times A_0 $ and $k' = \frac{k}{x}$.  We can then compute $A'_1$ : 

\begin{equation*}
   Non-Linearity'(k) =   Non-Linearity(k)
\end{equation*}

\begin{equation*}
    \frac{A'_1 \times (\frac{k}{x} +1)}{A_0 \times 2x} =   \frac{A_1\times(k+1)}{2 \times A_0}  
\end{equation*}

\begin{equation*}
    A'_1 = x \times A_1 \times \frac{k+1}{\frac{k}{x} +1} 
\end{equation*}

We approximate the ratio : $\frac{k+1}{\frac{k}{x} +1} \approx x$

\begin{equation*}
    A'_1 = x^2 \times A_1= A_1 \times (\frac{A'_0}{A_0})^2
\end{equation*}

We can finally write:

\begin{equation}
   \frac{A_1^\prime}{A_0^{\prime 2}}=\frac{A_1}{A_0^2} = \gamma
\end{equation}

\begin{figure}[ht!]
        \centering
        \includegraphics[width=11cm]{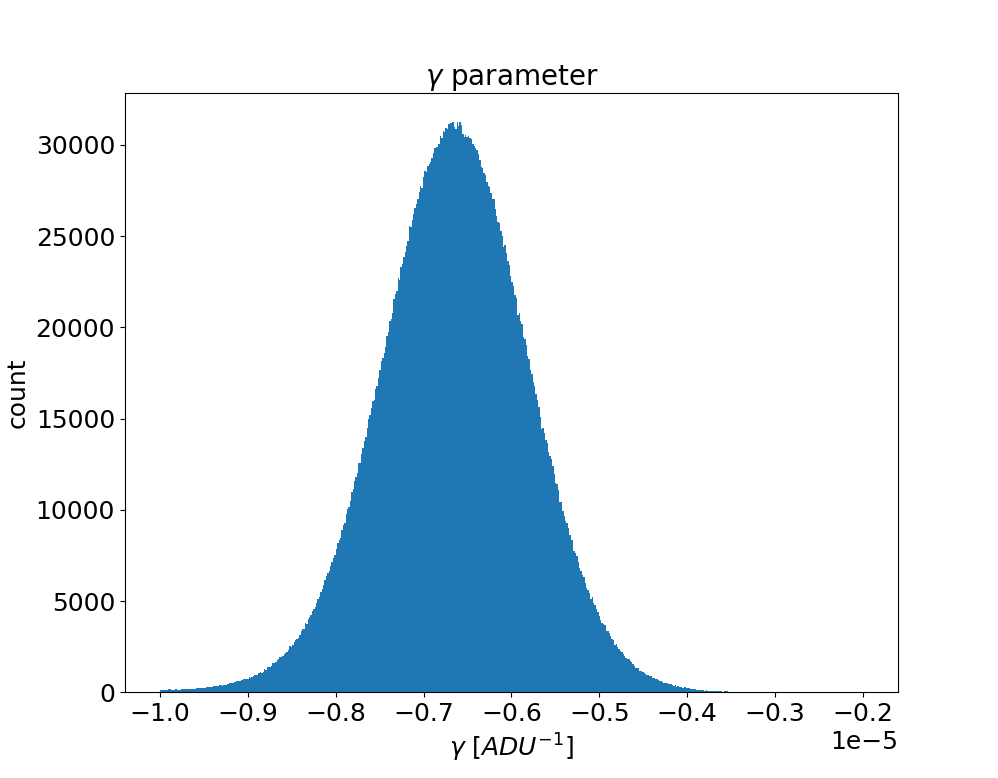}
        \caption{Histogram of the $\gamma$ coefficient, peaking around $-6.5\ 10^{-6}$.}
        \label{fig:gamma}
    \end{figure}

We here conclude that the non-linearity coefficient $\gamma$ defined by the ratio $\gamma = \frac{A_1}{A_0^2}$, is independent of the signal received. Figure \ref{fig:gamma} shows the histogram of this coefficient computed with calibration data. 

\end{document}